\documentclass[aps,pra,reprint,twocolumn,superscriptaddress,notitlepage]{revtex4-1}
\usepackage{amssymb,amsfonts,amsmath} 

\usepackage[pdftex]{graphicx} 
\usepackage{epstopdf}

\usepackage[utf8]{inputenc}
\usepackage[T1]{fontenc}
\usepackage{color}
\usepackage{setspace}
\usepackage{exscale}
\usepackage{transparent}
\usepackage[a4paper, total={6in, 9in}]{geometry}

\begin{document}


\title{Spin-asymmetric Josephson plasma oscillations}



\author{J.\ M.\ Kreula}
\affiliation{Clarendon Laboratory, University of Oxford, Parks Road, Oxford OX1 3PU, United Kingdom}

\author{G.\ Valtolina}
\affiliation{Istituto Nazionale di Ottica del Consiglio Nazionale delle Ricerche, 50019 Sesto Fiorentino, Italy}
\affiliation{European Laboratory for Nonlinear Spectroscopy, 50019 Sesto Fiorentino, Italy}

\author{P.\ T\"orm\"a}
\altaffiliation{Electronic address: paivi.torma@aalto.fi}
\affiliation{COMP Centre of Excellence, Department of Applied Physics, Aalto University, 00076 Aalto, Finland}

\begin{abstract}
The spin-asymmetric Josephson effect is a proposed quantum-coherent tunnelling phenomenon where Cooper-paired fermionic spin-$\frac{1}{2}$ particles, which are subjected to spin-dependent potentials across a Josephson junction, undergo frequency-synchronized alternating-current Josephson oscillations with spin-dependent amplitudes. Here, in line with present-day techniques in ultracold Fermi gas setups, we consider the regime of small Josephson oscillations and show that the Josephson plasma oscillation amplitude becomes spin-dependent in the presence of spin-dependent potentials while the Josephson plasma frequency is the same for both spin-components. Detecting these spin-dependent Josephson plasma oscillations provides a possible means to establish the yet-unobserved spin-asymmetric Josephson effect with ultracold Fermi gases using existing experimental tools.
\end{abstract}

\date{\today}

\maketitle

\section{Introduction}\label{sec:intro}

The Josephson effect~\cite{josephson1962possible} refers to the dynamics of macroscopic variables such as the relative phase and particle number in a bipartite quantum many-body system known as a Josephson junction. Possibly the best-known instance of this phenomenon is a supercurrent through a solid-state superconducting tunnel junction~\cite{anderson1963probable}. An analogous effect has also been demonstrated for example in superfluid $^3$He~\cite{backhaus1997direct} and $^4$He~\cite{sukhatme2001observation}, exciton polaritons~\cite{abbarchi2013macroscopic}, and ultracold atoms with both bosonic~\cite{PhysRevLett.95.010402,levy2007ac} and fermionic~\cite{valtolina2015josephson} species. Here, we consider the case of ultracold Fermi gases, although our results will be conceptually general and thus applicable also in other systems.

There has been earlier theoretical work on the Josephson effect in ultracold Fermi gases, see, e.g.,~\cite{PhysRevLett.99.040401,PhysRevB.77.144521,PhysRevA.79.033627,zou2014josephson}. However, the unprecedented control and tunability of parameters and individual degrees of freedom that are achievable in ultracold atomic gas setups~\cite{lewenstein2007ultracold,bloch2008many,theoryoffermi,chin2010feshbach,esslinger2010,bloch2012quantum,torma2014quantum} offer the possibility to consider going beyond the standard Josephson phenomenon. A case in point is the proposed spin-asymmetric Josephson effect~\cite{paraoanu2002josephson,miikkajose,kreula2014}.  In this scenario, the Cooper-paired fermionic particles are subjected to a spin-dependent potential $\delta_\sigma$ (here, $\sigma=\uparrow,\downarrow$) across the Josephson junction. As a result, the spins still display coherent Josephson oscillations with the same Josephson frequency for both components, but the  \emph{amplitude}, or the critical Josephson current~$I^C$, becomes \emph{spin-dependent}, i.e., the Josephson current for spin $\sigma$ has the form $I^J_\sigma(t)=I^C_{\sigma}(\delta_{\bar{\sigma}})\sin[(\delta_\uparrow+\delta_\downarrow)t + \varphi]$, where $t$ denotes time and  $\varphi$ is the initial phase difference.  We set $\hbar=1$ throughout this work. Note that the critical current depends only on the potential of the opposite spin, $\delta_{\bar{\sigma}}$, but the Josephson frequency, $\omega_J=\delta_\uparrow+\delta_\downarrow$, is spin-symmetric. 

The physical origin of this rather surprising result can be elucidated by considering the dynamics of a single Cooper pair across the Josephson junction in the presence of spin-dependent potentials and tunnelling couplings $\Omega_\sigma$, as shown in~\cite{miikkajose}. The relevant initial state is a superposition of the paired states on the left and right-hand sides of the junction, given by $\alpha_0 |\uparrow \downarrow\rangle_L |\emptyset \rangle_R + \beta_0 |\emptyset \rangle_L |\uparrow \downarrow\rangle_R$, where $\alpha_0$ and $\beta_0$ are complex numbers with $|\alpha_0|^2 + |\beta_0|^2=1$. The broken-pair `intermediate' states, ${|\uparrow \rangle_L |\downarrow \rangle_R}$ and ${| \downarrow\rangle_L |\uparrow \rangle_R}$, are required to describe the tunnelling processes. If no spin-dependent potentials are present, these intermediate states have the same eigenenergy. However, the degeneracy is lifted by the spin-dependent potentials. It turns out that this is the key to understanding the origin of the phenomenon.

\begin{figure*}
\centerline{\includegraphics[scale=0.7]{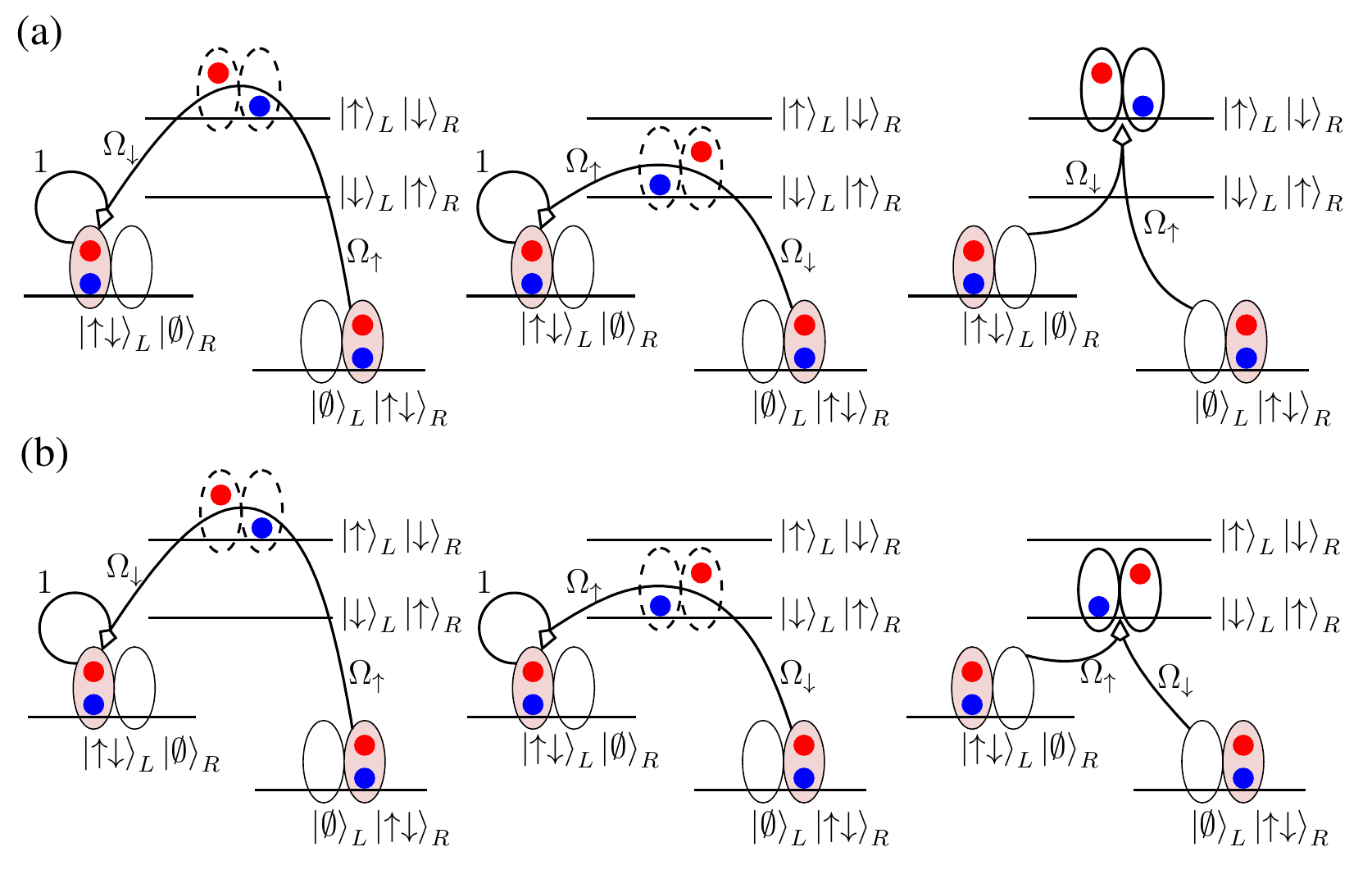}}
\caption{(Color online) Origin of the spin-asymmetric critical Josephson currents. Considering the dynamics of a single Cooper pair across a Josephson junction with spin-dependent potentials, the Josephson current results from a superposition of the paired states, $|\uparrow \downarrow \rangle_L |\emptyset \rangle_R$ and $|\emptyset \rangle_L |\uparrow \downarrow\rangle_R$. Here, we show the tunnelling processes that contribute to (a) $I_\uparrow^J$ and to (b) $I_\downarrow^J$.  The three processes are the following. First, there is a pair-tunnelling contribution via the intermediate state $|\uparrow \rangle_L |\downarrow \rangle_R$ (left panel). Second, there is another pair-tunnelling contribution via the other intermediate state $|\downarrow \rangle_L |\uparrow \rangle_R$ (middle panel). We have included the loop with label 1 to remind that these usual Josephson processes describe the interference between the tunnelled pair and the initial population (indicated by the label 1) of the state $|\uparrow \downarrow \rangle_L |\emptyset \rangle_R$. The pair-tunnelling processes are the same for both spin components. Finally, we find that there is also a virtual \emph{single-particle interference} contribution (right panel). The single-particle interference term is different for the spin-$\uparrow$ and spin-$\downarrow$ components due to the presence of the spin-dependent potential which lifts the energy degeneracy of the intermediate states. This causes the spin-asymmetric Josephson effect.}
\label{fig:joseorigin}
\end{figure*}

We proceed to explain the different tunnelling processes constituting the Josephson current. In the simplified system of a single Cooper pair, which is analogous to a two-site Hubbard model, time-dependent perturbation theory to second order in $\Omega_\sigma$ yields the Josephson current for spin~$\sigma$ as
\begin{align}\label{eq:Ijtoy}
I^J_\sigma(t)=&2\Omega_\uparrow\Omega_\downarrow |\alpha_0 \beta_0| \left(M_{\rm pair} + M_{\rm single}^{\sigma} \right) \nonumber \\ & \times \sin\left[\left({\delta}_{\uparrow}+{\delta}_{\downarrow}\right)t + \varphi \right]. 
\end{align}
For details of the calculation, see~\cite{miikkajose}. Here, $M_{\rm pair}=\frac{1}{U+{\delta}_\uparrow}+\frac{1}{U+{\delta}_\downarrow}$, where $U$ is the interaction between the spin components. This term results from second-order tunnelling processes starting from the state ${|\emptyset\rangle_L |\uparrow \downarrow\rangle_R}$  and ending in the state ${|\uparrow \downarrow\rangle_L |\emptyset \rangle_R}$ via either the state ${|\uparrow \rangle_L |\downarrow \rangle_R}$ or the state ${| \downarrow\rangle_L |\uparrow \rangle_R}$. Thus, $M_{\rm pair}$ describes the usual pair interference process that is symmetric with respect to ${\delta}_\uparrow$ and ${\delta}_\downarrow$. It turns out that there is also a contribution from two first-order processes that break the spin-symmetry. These processes yield the terms ${M_{\rm single}^{\uparrow}=\frac{1}{U-{\delta}_\downarrow}-\frac{1}{U+{\delta}_\uparrow}}$ and ${M_{\rm single}^{\downarrow}=\frac{1}{U-{\delta}_\uparrow}-\frac{1}{U+{\delta}_\downarrow}}$ which are different for the two spin components. The term $M_{\rm single}^{\uparrow}$ is the result of the interference of the virtual broken-pair tunnelling processes ${  |\uparrow \downarrow \rangle_L |\emptyset \rangle_R \rightarrow |\uparrow \rangle_L |\downarrow \rangle_R}$ and ${|\emptyset \rangle_L |\uparrow \downarrow\rangle_R \rightarrow |\uparrow \rangle_L |\downarrow \rangle_R}$, while $M_{\rm single}^{\downarrow}$ emerges from the interference of $ { |\uparrow \downarrow \rangle_L |\emptyset \rangle_R \rightarrow |\downarrow \rangle_L |\uparrow \rangle_R}$ and $|\emptyset \rangle_L |\uparrow \downarrow\rangle_R \rightarrow |\downarrow \rangle_L |\uparrow \rangle_R$. Since the energy degeneracy of the intermediate states is lifted by the presence of the spin-dependent potentials, the virtual broken-pair tunnelling processes contribute asymmetrically to the Josephson current, and thus produce the spin-asymmetric Josephson effect. However, note that the single-particle processes are present also in the standard symmetric case, $\delta_\uparrow=\delta_\downarrow$.  We also emphasize that these virtual broken-pair tunnelling processes do not refer to the cosine-term (the `quasiparticle interference term') of the Josephson effect which involves actual single-particle transitions and vanishes at zero temperature for potentials smaller than the excitation gap 2$\Delta$. The different interference processes contributing to the Josephson current are depicted in Fig.~\ref{fig:joseorigin}. 

The Josephson junction with spin-dependent potentials has similarities to ferromagnetic Josephson junctions~\cite{PhysRevLett.86.2427,bergeret2001enhancement,keizer2006spin,PhysRevLett.104.137002,PhysRevLett.110.117003,meng2013long}. Perhaps most notably, the tunable critical supercurrent in SFIFS junctions~\cite{bergeret2001enhancement} (here, S stands for superconductor, I for insulator, and F for ferromagnet) can be explained by the spin-asymmetric Josephson effect in the direct-current (dc) limit at zero temperature~\cite{kreula2014}. We, however, emphasize that in the spin-asymmetric Josephson effect the spin-dependent potentials create the asymmetry, and the barrier separating the superconductors can be just an insulator without a spin-active coupling, unlike in the case of a ferromagnetic barrier. In fact, the spin-asymmetric Josephson effect could possibly be realized in a solid-state SIS junction with two superconductors that have different Zeeman splittings for the two spin states in the presence of a magnetic field~\cite{meservey1994spin}. There is also no immediate connection between our single-particle interference terms and Andreev reflections in weak links, since Andreev reflections can take place even without Josephson effects, whereas our virtual single-particle interferences are always inherent to the coherent Josephson current regardless of the type of the junction. Moreover, our single-particle interference term vanishes in the dc Josephson effect (see the discussion after Eq.~\eqref{eq:Ijtoy}), while Andreev reflections and bound states can be relevant also in the dc limit~\cite{golubov2004current}. Finally, we point out that the spin-asymmetric Josephson effect occurs when the pairing is of singlet-type, and no triplet-pairing is required.

The spin-asymmetric Josephson effect has been predicted to take place between the hyperfine levels of a four-component superfluid Fermi gas with radio-frequency (RF) field induced transitions and in two-component superfluids in spin-dependent double wells~\cite{paraoanu2002josephson,miikkajose}. It has also been proposed to occur within a single superfluid between the odd and even sites of a spin-dependent superlattice~\cite{kreula2014}. 

Recently~\cite{valtolina2015josephson}, the observation of Josephson plasma oscillations throughout the BCS--BEC crossover, where BCS stands for Bardeen--Cooper--Schrieffer and BEC for Bose--Einstein condenstate, has been reported in ultracold Fermi gases. Here, motivated by these experimental advances in Josephson dynamics, we consider the possibility to observe the spin-asymmetric Josephson effect via \emph{spin-dependent plasma oscillations} in a superfluid Fermi gas forming a Josephson junction with a spin-dependent potential. All the required experimental tools that we will present have already been demonstrated with ultracold atoms. Thus, the arrangement that we propose can be realized with existing techniques in ultracold atom setups. The plasma oscillation regime corresponds to a relative number difference on the order of a few per cent across the junction~\cite{zou2014josephson,valtolina2015josephson}. This limits the possible values of the spin-dependent potentials (see Section~\ref{sec:plasmares}).

In comparison to earlier studies on the spin-asymmetric Josephson effect, we point out that Josephson plasma oscillations differ from the full alternating-current (ac) oscillations in that they refer to the solution of linearized Josephson equations, similar to the equations of motion of a classical pendulum, with the small-angle approximation for the Josephson phase. This leads to small amplitude oscillations of the number density with a plasma frequency that is different from the Josephson frequency given by the potential difference across the junction. Full ac Josephson oscillations have not yet been observed in ultracold Fermi gases.

This work is organized as follows. In Section~\ref{sec:setup} we describe our setup with especially ultracold Fermi gases in mind. We present our results in Section~\ref{sec:plasma}. Finally, we end with a summary and discussion in Section~\ref{sec:summary} before describing some calculation details in the Appendix.

\section{Setup}\label{sec:setup}
We give the description of the setup for the spin-asymmetric Josephson effect in terms of ultracold atoms. However, we emphasize that the arrangement is conceptually general and as such not only restricted to ultracold atoms, and thus other systems could also be considered.

Our setup consists of a two-component superfluid atomic Fermi gas (e.g., $^6$Li) divided into two weakly connected reservoirs, $L$ and $R$, with chemical potentials $\mu_L$ and $\mu_R$, respectively. This setup forms an effective Josephson junction. The arrangement is similar to the ones used in the observation of Josephson dynamics in $^6$Li~\cite{valtolina2015josephson} and in quantum transport experiments with $^6$Li~\cite{brantut2012conduction,stadler2012observing,brantut2013thermoelectric,krinner2015observation,husmann2015connecting,krinner2016mapping}.

In the absence of couplings between the reservoirs and additional potentials, the system is described with the Hamiltonian
\begin{align}
&\hat{H}_0\nonumber \\ &= \int  d{\bf r}\, \sum_{\sigma}\sum_{j=L,R} \Big[ \hat{\psi}_{\sigma,j}^{\dagger}({\bf r})\left(-\frac{\nabla^2}{2m}-\mu_j \right)\hat{\psi}_{\sigma,j}({\bf r})  \Big]\nonumber \\
&+\int  d{\bf r_1} \int  d{\bf r_2}\, \sum_{j=L,R} U({\bf r_1,r_2})\nonumber \\ & \times \Big[ \hat{\psi}^{\dagger}_{\uparrow,j}({\bf r_1})\hat{\psi}^{\dagger}_{\downarrow,j}({\bf r_2})\hat{\psi}_{\downarrow,j}({\bf r_2})\hat{\psi}_{\uparrow,j}({\bf r_1}) \Big],
\end{align}
where $\hat{\psi}_{\sigma,L/R}^{\dagger}({\bf r})$ ($\hat{\psi}_{\sigma,L/R}({\bf r})$) is a field operator that creates (annihilates) a spin $\sigma$ fermion at position ${\bf r}$ on the left/right hand side of the junction. Consistent with dilute atomic Fermi gases, we assume an attractive contact interaction, $U({\bf r_1,r_2})=g\delta({\bf r_1-r_2})$, between the spin $\uparrow$ and $\downarrow$ particles on both sides of the junction. Here, $\delta({\bf r})$ is the Dirac delta function in three dimensions (3D) and ${g=\frac{4\pi }{m}a_s}<0$, where $m$ is the mass of the particles and $a_s$ is the 3D $s$-wave scattering length. As is customary with ultracold fermions, we give the interaction in terms of the dimensionless parameter $k_Fa_s$, where $k_F$ is the Fermi momentum, as $g=\frac{8}{3\pi}k_Fa_s \frac{E_F}{n}$. Here, $E_F$ denotes the Fermi energy and $n$ the particle number density.

\begin{figure}[t!]
\centerline{\includegraphics[scale=0.42]{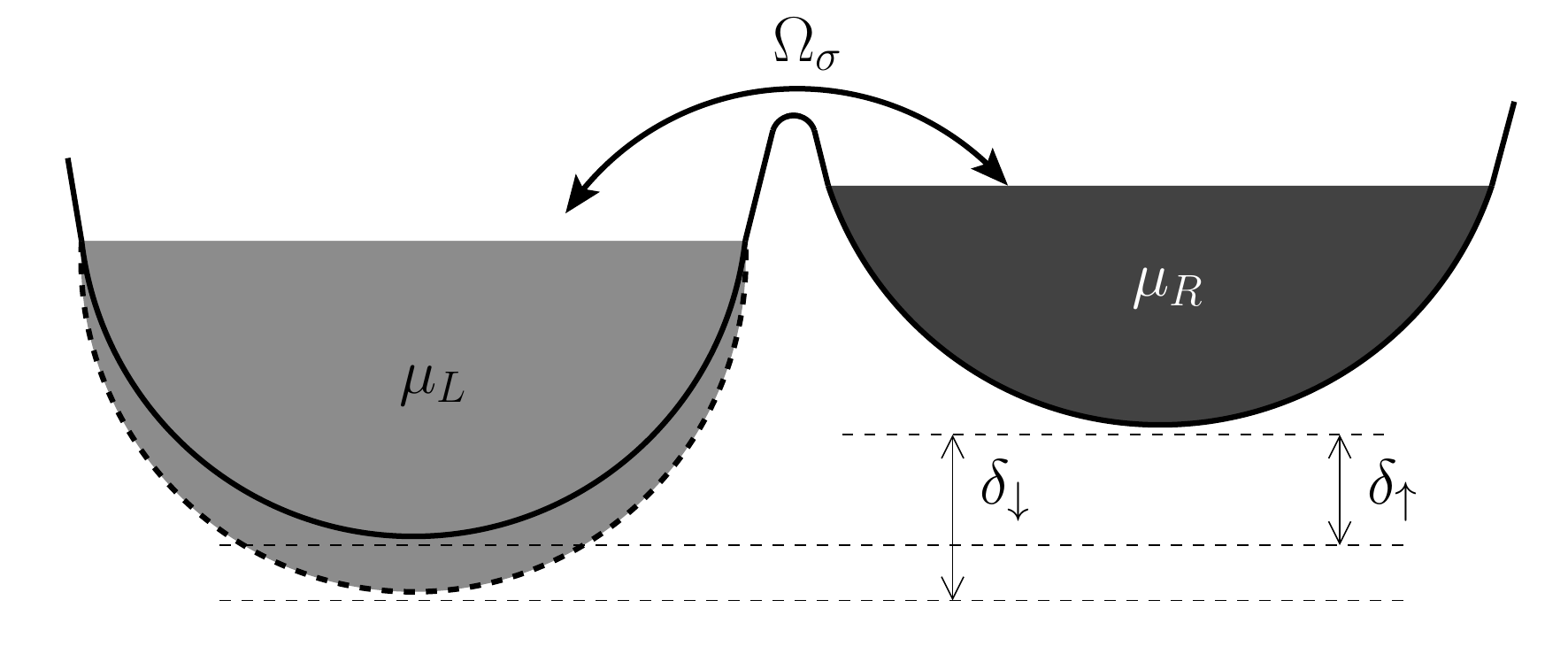}}
\caption{Schematic of a spin-dependent Josephson junction. A superfluid Fermi gas is divided into two reservoirs denoted by $L$ and $R$. The reservoirs are connected via the weak tunnelling coupling $\Omega_{\sigma}$, which induces Josephson oscillations across the junction. Additionally, a spin-dependent potential $\delta_\sigma$ is applied across the junction, which creates a spin-asymmetry in the Josephson current.}
\label{fig:dw}
\end{figure}

To induce Josephson currents in the system, we assume that at time $t=0^+$ the two reservoirs become weakly coupled via the Hamiltonian
\begin{align}\label{eq:couplHam}
\hat{H}_{\Omega}=\sum_\sigma \Omega_{\sigma} \int  d{\bf r}\, \hat{\psi}_{\sigma,L}^{\dagger}({\bf r})\hat{\psi}_{\sigma,R}({\bf r}) + {\rm H.c.},
\end{align}
where $\Omega_{\sigma}$ is the coupling strength. Moreover, to create the spin-asymmetry in the currents, an additional spin-dependent potential $\delta_\sigma$ is applied across the junction at time $t=0^+$. This is described with the Hamiltonian
\begin{align}
\hat{H}_{\delta}=&\int  d{\bf r}\, \sum_{\sigma} \Big[ \left(\mu_L - \frac{\delta_\sigma}{2} \right)\hat{\psi}_{\sigma,L}^{\dagger}({\bf r})\hat{\psi}_{\sigma,L}({\bf r})  \nonumber\\ &+\left(\mu_R + \frac{\delta_\sigma}{2} \right) \hat{\psi}_{\sigma,R}^{\dagger}({\bf r})\hat{\psi}_{\sigma,R}({\bf r}) \Big].
\end{align}
For the reason why the chemical potentials are included also in this Hamiltonian, see Chapter 10.4.1 in Ref.~\cite{torma2014quantum}. Alternatively, as demonstrated in~\cite{krinner2016mapping}, one can consider spin-dependent chemical potentials, $\mu_{L,\sigma} \neq \mu_{R,\sigma}$, switched on at time $t=0^+$. The total Hamiltonian of the system thus reads
\begin{align}\label{eq:totalham}
\hat{H}=\hat{H}_0+\hat{H}_{\Omega}+\hat{H}_{\delta}.
\end{align}
See Fig.~\ref{fig:dw} for a schematic illustration of the system.

We now describe how the required spin-dependent potential could be achieved experimentally in two different ways by utilizing spin-dependent interactions. In an ultracold Fermi gas setup, the spin $\sigma$ corresponds to, e.g., the lowest two hyperfine levels of $^6$Li, $|1\rangle$ and $|2\rangle$, with, e.g., $|\downarrow \rangle=|1 \rangle$ and $|\uparrow \rangle=|2 \rangle$, which features a Feshbach resonance at a magnetic field strength of 832 G.

Our first suggested implementation exploits a third spin-component, e.g., atoms transferred via an RF pulse to the third lowest hyperfine level of $^6$Li, denoted by $|3\rangle$, introduced on one side of the junction. On the BCS side of the 1-2 Feshbach resonance, i.e., for magnetic fields above 832 G, the atoms in the state $|3\rangle$ interact differently with the atoms in $|1 \rangle$ and $|2\rangle$ due to the different positions of the respective pairwise Feshbach resonances~\cite{Zurn2012}. This allows utilizing the mean-field Hartree shift to create a potential difference between the atoms in the states $|1 \rangle$ and $|2\rangle$, given by $\Delta\delta_{12}= \frac{4\pi}{m}n_3(a_{13}-a_{23})$, where $n_3$ is the number density of atoms in state $|3\rangle$, and $a_{13}$ ($a_{23}$) is the scattering length for collisions between atoms in states $|1\rangle$ ($|2\rangle$) and $|3\rangle$ (see, e.g.,~\cite{regal2003measurement}). In this magnetic field regime, $\Delta \delta_{12}$ is significant even for $n_3\simeq 10^{10}$ cm$^{-3}$. Such a low density is required to allow a lifetime on the order of a few hundred milliseconds against three-body recombinations in three-component Fermi gases, in accordance with~\cite{Huckans2009, Williams2009, Ottenstein2008}.

Our second proposal is based on the recent experimental realization of Bose--Fermi superfluid mixtures~\cite{Ferrier2014}, which has been achieved even in systems with a large mass-imbalance~\cite{Yao2016, Roy2016}. In particular, we suggest to create a Bose--Fermi superfluid mixture where the bosonic atoms are either $^{87}$Rb or $^{133}$Cs, both of which feature broad Feshbach resonances on the BCS side of the 1-2 $^6$Li superfluid. This allows the tuning of the interaction with respect to only one $^6$Li spin component~\cite{Deh2008,Repp2013} and again creates a spin-dependent mean-field Hartree shift, which for a Bose-Fermi mixture can be large also away from the center of the resonance~\cite{Zaccanti2006}. 

In both of the proposed schemes, as the two spin components $|1 \rangle$ and $|2\rangle$ tunnel through the Josephson junction, the difference in the mean-field shifts, $\Delta \delta_{12}$, creates the required spin-dependent potential difference $\delta_\sigma$ across the junction. Since the scattering lengths can be tuned and the number density can be controlled, this spin-dependent potential difference can be varied as well.

\section{Results}\label{sec:plasma}
\subsection{Spin-asymmetric Josephson currents}
We are interested in calculating the Josephson currents in our system and take the spin-dependent potentials $\delta_\sigma$ to be smaller than the excitation gap 2$\Delta$. The setup presented in Section~\ref{sec:setup} is mathematically analogous to that considered in our previous work on the spin-asymmetric Josephson effect in a four-component Fermi gas~\cite{miikkajose}. Thus, the same calculations for the Josephson currents presented in~\cite{miikkajose}  and the Supplemental Material therein using the BCS mean-field approach, linear response theory, and the Kadanoff--Baym formalism~\cite{baym1961conservation,kadanoff1962quantum} are also applicable in this setup. Here we show only the result. We present an outline of the calculations in the Appendix.

The Josephson current for spin $\sigma$ is given by
\begin{align}\label{eq:josecurr}
I^J_{\sigma}(t)=-I^C_{\sigma}(-\tilde{\delta}_{\bar{\sigma}})\sin\left[\left (\tilde{\delta}_\uparrow + \tilde{\delta}_\downarrow\right)t - \varphi \right],
\end{align}
where $\tilde{\delta}_\sigma=\mu_L-\mu_R-\delta_\sigma$. Here the critical Josephson current reads
\begin{align}\label{eq:criticalc}
I^C_{\sigma}(-\tilde{\delta}_{\bar{\sigma}})=2\left| \Omega_\uparrow \Omega_\downarrow \Pi_{\mathcal{F}} ({\bf p=0},-\tilde{\delta}_{\bar{\sigma}} + i 0^+ ) \right|,
\end{align}
where
\begin{align}\label{eq:Fbubble}
\Pi_{\mathcal{F}} ({\bf p}, i\omega_n )=& \frac{1}{\beta V} \sum_{{\bf q},i\omega_m} \mathcal{F}_L({\bf q},i\omega_m) \nonumber \\ & \times \mathcal{F}^{\dagger}_R({\bf q-p},i\omega_m-i\omega_n).
\end{align}
In this expression, $\omega_n$ and $\omega_m$ denote fermionic Matsubara frequencies,  $V$ is the volume, and $\beta=\frac{1}{k_B T}$, where $k_B$ is the Boltzmann constant and $T$ is temperature. Furthermore, $\mathcal{F}_L$ ($\mathcal{F}_R$) is the anomalous BCS--Nambu--Gor'kov Green function describing Cooper pairing correlations on the left (right) hand side of the junction. 
The anomalous Green function is given in Matsubara space by
\begin{align}\label{eq:anomalousG}
\mathcal{F}({\bf q},i\omega_m)=\frac{\Delta}{(i\omega_m)^2-E_{\bf q}^2},
\end{align}
where $E_{\bf q}=\sqrt{\xi_{\bf q}^2+\Delta^2}$. Here,  $\xi_{\bf q}$ is the kinetic energy of momentum state ${\bf q}$ given relative to the chemical potential, i.e, $\xi_{\bf q}=\epsilon_{\bf q}-\mu$, with $\epsilon_{\bf q}=|{\bf q}|^2/2m$. 

We point out again that in Eq.~\eqref{eq:josecurr} the critical current for the spin $\sigma$ component depends only on the potential $\tilde{\delta}_{\bar{\sigma}}$ of the opposite spin, while the Josephson frequency, $\omega_J=\tilde{\delta}_\uparrow + \tilde{\delta}_\downarrow$, is the same for both spin components. This is the spin-asymmetric Josephson effect. 

Note that even though these fully coherent Josephson oscillations are spin-dependent, there is no total equilibrium spin-imbalance in our system. This suppresses the possibility for equilibrium phase separation, and for exotic Fulde--Ferrell--Larkin--Ovchinnikov-type pairing that is furthermore unlikely to occur in the 3D case considered here~\cite{zwierlein2006fermionic,partridge2006pairing,kinnunen2006strongly}.

\subsection{Spin-asymmetric plasma oscillations}\label{sec:plasmares}

Motivated by the experiment in~\cite{valtolina2015josephson}, we now ask how the small-amplitude Josephson plasma oscillations are affected by the presence of spin-dependent potentials. Detecting spin-asymmetric Josephson plasma oscillations offers an alternative and perhaps an experimentally more feasible means to establish the spin-asymmetric Josephson effect with present-day techniques. We leave the non-linear self-trapping regime~\cite{PhysRevA.59.620,PhysRevLett.95.010402,abbarchi2013macroscopic} in the presence of spin-dependent potentials for future work.

To begin the analysis, we introduce a spin-dependent number difference parameter $\Delta N^J_\sigma= \frac{1}{2}(\langle \hat{N}^J_{\sigma,L} \rangle - \langle \hat{N}^J_{\sigma,R} \rangle)=\frac{1}{2}(N^J_{\sigma,L}-N^J_{\sigma,R})$, akin to the bosonic case~\cite{pitaevskii2003bose}. Here, $\hat{N}^J_{\sigma,L}$  ($\hat{N}^J_{\sigma,R}$) denotes the number  operator for spin $\sigma$ particles on the left (right) reservoir that belong to the Fermi condensate and can thus contribute to the Josephson current. Using Eq.~\eqref{eq:josecurr} and the fact that $\partial_t N^J_{\sigma,L}=-\partial_t N^J_{\sigma,R}$, we find that the dynamics of $\Delta N^J_\sigma$ is obtained from
\begin{align}
\frac{\partial \left(\Delta N^J_\sigma \right)}{\partial t}=-I^C_\sigma(-\tilde{\delta}_{\bar{\sigma}})\sin \Phi^J(t).
\end{align}
The Josephson phase $\Phi^J(t)$ obeys the equation of motion
\begin{align}\label{eq:phi}
\frac{\partial \Phi^J(t)}{\partial t} = \tilde{\delta}_\uparrow + \tilde{\delta}_\downarrow= 2(\mu_L - \mu_R) - {\delta}_\uparrow - {\delta}_\downarrow.
\end{align}

The spin-dependent critical Josephson current implies spin-dependent number oscillations also in the plasma oscillation regime. Therefore also the chemical potential undergoes small spin-dependent dynamics. We can write Eq.~\eqref{eq:phi} as
\begin{align}
\frac{\partial \Phi^J(t)}{\partial t} = E^{\uparrow}_{Ch} \Delta N^J_{\uparrow} + E^{\downarrow}_{Ch} \Delta N^J_{\downarrow} - {\delta}_\uparrow - {\delta}_\downarrow,
\end{align}
where we have introduced the spin-dependent charging energy $E^{\sigma}_{Ch}=2\frac{d\mu_{\sigma,L}}{dN^J_{\sigma,L}}$ which is evaluated at $N^J_{\sigma,L}=N^J_{\sigma,R}=N^J_{\sigma}/2$~\cite{pitaevskii2003bose}.

\begin{figure*}[t!]
\centerline{
\includegraphics[scale=0.257]{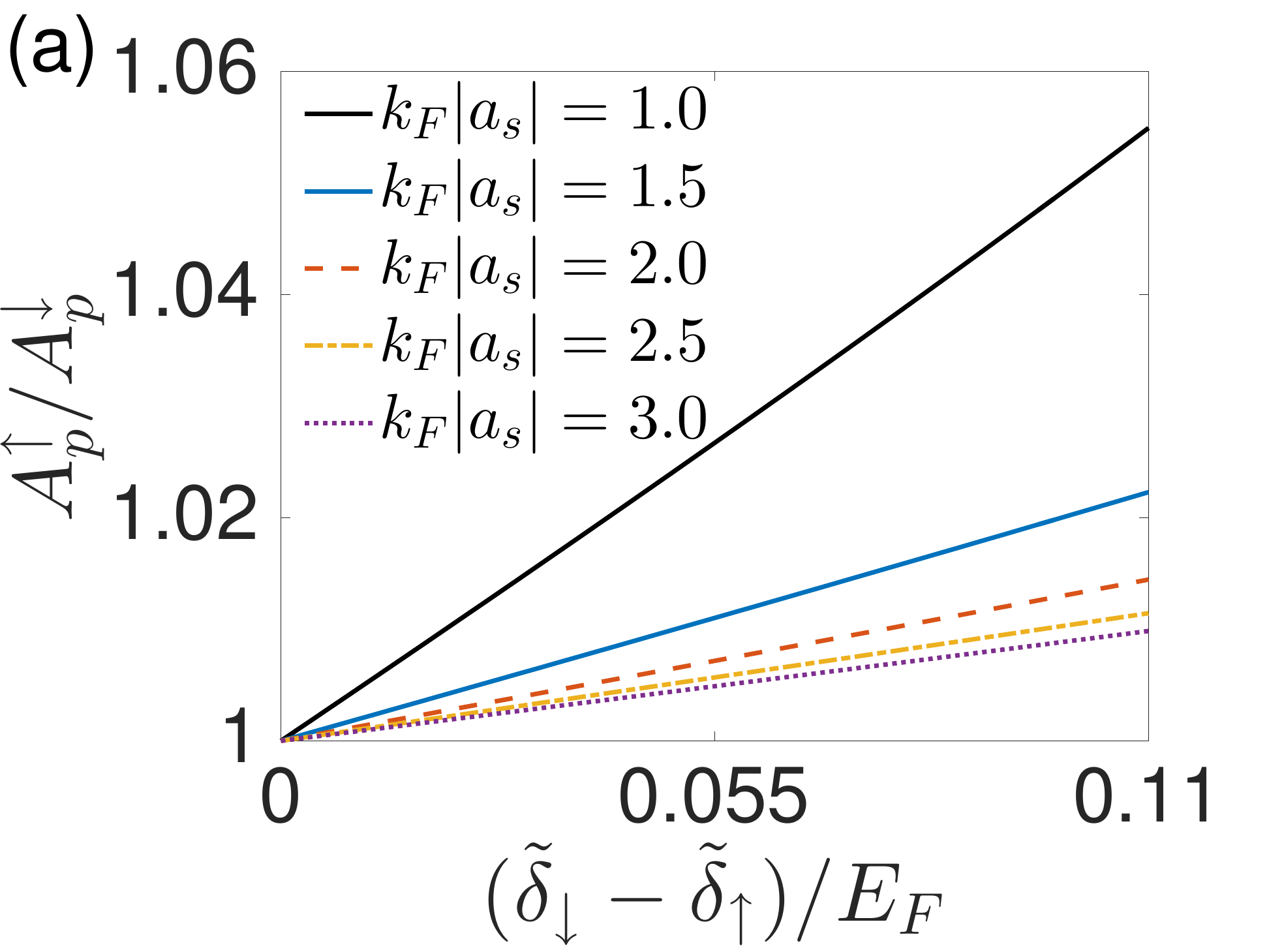}
\includegraphics[scale=0.257]{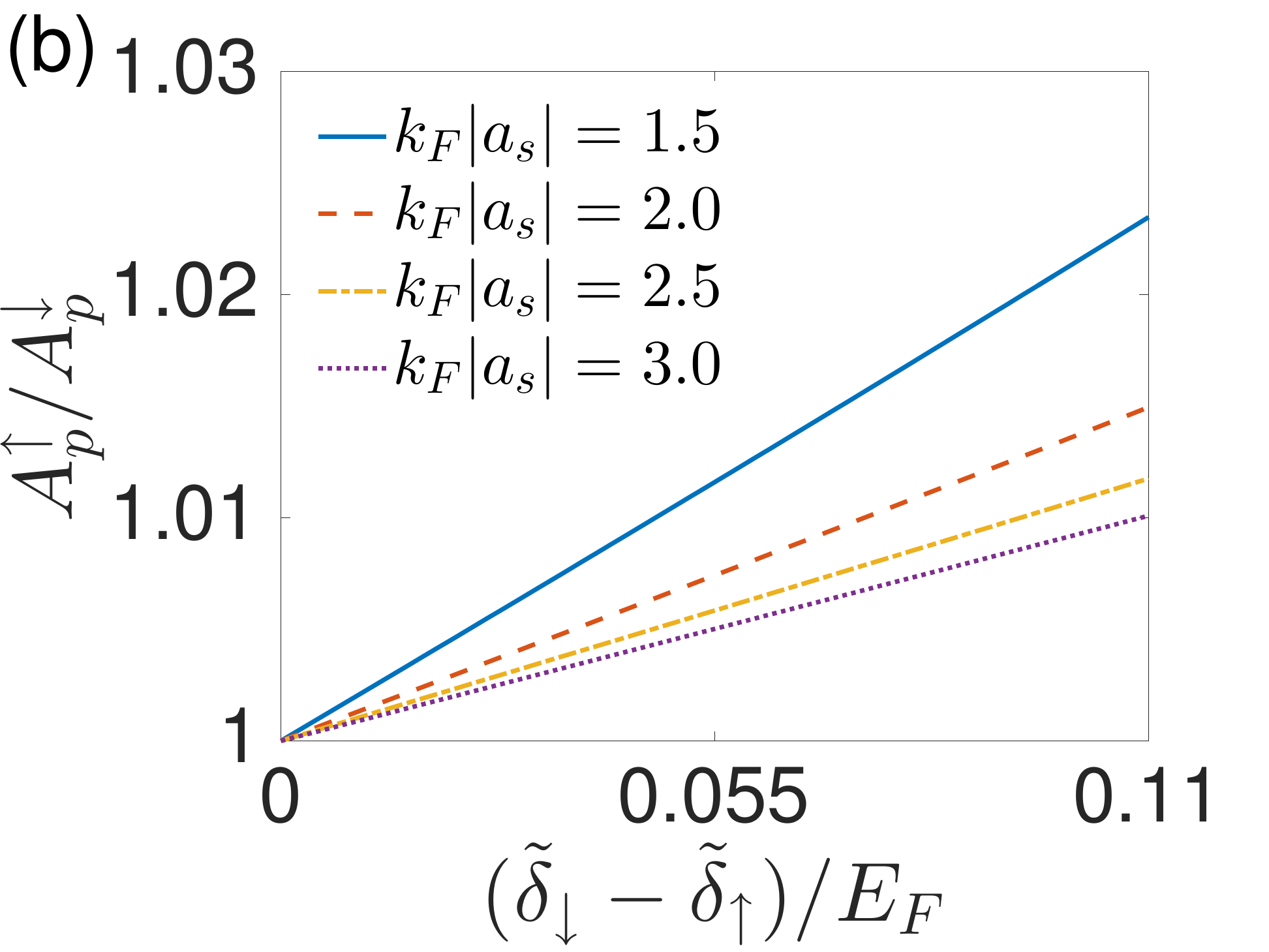}
\includegraphics[scale=0.257]{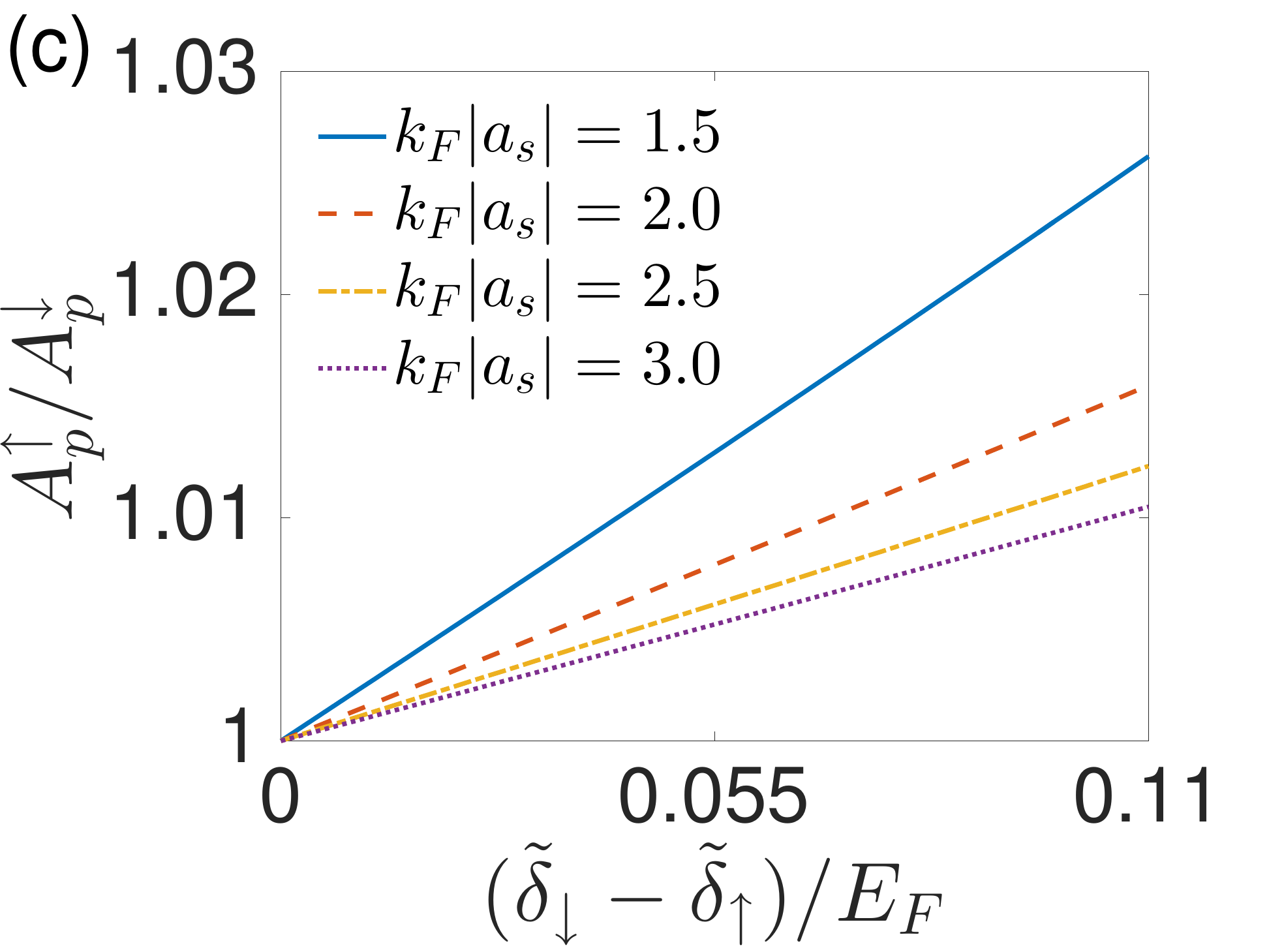}
}
\caption{(Color online) Asymmetry in the spin-dependent plasma oscillation amplitudes as a function of the difference in the spin-dependent potentials for Josephson frequency $\omega_J=0.11E_F$ and interaction strength $k_Fa_s=-1.0$ (black solid curve, only in (a)), $k_Fa_s=-1.5$ (blue solid curve), $k_Fa_s=-2.0$ (red dashed curve), $k_Fa_s=-2.5$ (yellow dash dotted curve), and $k_Fa_s=-3.0$ (purple dotted curve). The temperature is (a) $T=0.05T_F$, (b) $T=0.07T_F$, and (c) $T=0.09T_F$. The temperature regime is the same as in the experiment in~\cite{valtolina2015josephson}. In (a), the $k_Fa_s=-1.0$ curve is included as the reference line to the unitary Fermi gas regime.}
\label{fig:asymplasma}
\end{figure*}

For times much shorter than the inverse Josephson frequency but long enough to observe plasma oscillations, we have~$\sin \Phi^J(t) \approx \Phi^J(t)$, and we obtain the coupled differential equations 
\begin{align}
\frac{\partial^2  (\Delta N^J_\uparrow )}{\partial t^2}& = -I^C_\uparrow(-\tilde{\delta}_{\downarrow}) \nonumber \\ & \times \left( E^{\uparrow}_{Ch} \Delta N^J_\uparrow +E^{\downarrow}_{Ch} \Delta N^J_\downarrow  -  {\delta}_\uparrow - {\delta}_\downarrow \right),
\end{align}
and
\begin{align}
\frac{\partial^2  (\Delta N^J_\downarrow )}{\partial t^2}& = -I^C_{\downarrow}(-\tilde{\delta}_{\uparrow}) \nonumber \\ & \times \left( E^{\uparrow}_{Ch}  \Delta N^J_\uparrow +E^{\downarrow}_{Ch} \Delta N^J_\downarrow  -  {\delta}_\uparrow - {\delta}_\downarrow \right).
\end{align}
The solution for spin $\sigma$ has the form
\begin{align}
\Delta N^J_\sigma(t)=A^\sigma_p \sin (\omega_p t)+\Delta N^J_\sigma(0),
\end{align}
where the Josephson plasma frequency is given by ${\omega_p = \sqrt{E^{\uparrow}_{Ch} I^C_\uparrow + E^{\downarrow}_{Ch} I^C_{\downarrow}}}$. Note that in the spin-symmetric case, we have the standard formula $\omega_p = \sqrt{E_{Ch} E_J}$, where $E_J=I^C_\uparrow + I^C_{\downarrow} = I^C$ is the Josephson energy~\cite{pitaevskii2003bose,valtolina2015josephson} (recall that $\hbar=1$). However, in general we expect the plasma frequency in the presence of spin-dependent potentials to differ from the standard case, as can be seen for example with the toy calculation presented in Section~\ref{sec:intro}.  

Similar to the full ac spin-asymmetric Josephson effect, we therefore find that the system undergoes frequency-synchronized Josephson plasma oscillations with \emph{a spin-dependent amplitude}. The asymmetry in the amplitudes follows the relation
\begin{align}\label{eq:asymplasma}
\frac{A^\uparrow_p}{A^\downarrow_p}=\frac{I^C_\uparrow(-\tilde{\delta}_{\downarrow})}{I^C_\downarrow(-\tilde{\delta}_{\uparrow})}.
\end{align}
Equation~\eqref{eq:asymplasma} is the main result of this paper.

The asymmetry given by Eq.~\eqref{eq:asymplasma} is limited by the requirement that the oscillations in the number density between the two reservoirs must be small. For the plasma oscillation approximation to be valid, the relative number difference, $z=\frac{N^J_L-N^J_R}{N^J_L+N^J_R}$, has to be on the order of a few per cent~\cite{zou2014josephson,valtolina2015josephson}. This gives an upper bound to the Josephson frequency, $\omega_J={\tilde{\delta}_{\downarrow}+\tilde{\delta}_{\uparrow}}$, in the plasma oscillation regime. The maximal Josephson frequency then determines how much the asymmetry in the plasma oscillation amplitudes can be tuned and how large asymmetries can be obtained.

To give an estimate of the upper bound for $\omega_J$, we assume for simplicity that there are no spin-dependent potentials across the junction. The Josephson dynamics is then induced only by the difference in the chemical potentials, with $\omega_J=2(\mu_L - \mu_R)$. We want to express $\omega_J$ in terms of the relative number difference $z$, whose values corresponding to the plasma oscillation regime are known~\cite{zou2014josephson,valtolina2015josephson}. Using the chemical potential for a non-interacting trapped gas, $\mu \propto N^{1/3}$, we find that the relative difference in the chemical potentials and in the particle numbers obey the relation $\eta=(\mu_L - \mu_R) /(\mu_L + \mu_R) = z/3$, which yields the Josephson frequency as $\omega_J=4\eta \mu=\frac{4}{3}z \mu$, where we have denoted $\mu=\mu_L \approx \mu_R$. With the equation $\eta=z/3$, we note that the relative chemical potential difference $\eta$ across the junction can only be 1\% with the parameters in~\cite{valtolina2015josephson}, i.e., $z=3\%$ (with $10^5$ atoms per spin state) and a barrier height of $1.2\pm0.1E_F$ between the reservoirs. In~\cite{zou2014josephson}, the critical value for $z$ was found to be on the order of 9\% for a barrier height of $5E_F$, and thus $\eta\approx 3\%$ in this case. In what follows, we use $\eta=3\%$ as the maximum relative difference in the chemical potentials to estimate the maximal Josephson frequency.

To get $\omega_J$, we need a value for the chemical potential $\mu$. In our simple case of a homogeneous Fermi gas, BCS mean-field theory with the attractive interaction strength $k_F|a_s|$ between $1.0$ and $3.0$ and the temperature $T$ between $0.05T_F$ and $0.09T_F$, where $T_F$ is the Fermi temperature, yields a chemical potential between approximately $0.81E_F$ and $0.96E_F$. This implies that the maximal Josephson frequency corresponding to the plasma oscillation regime is around $\omega_J^{\max} \approx 0.11E_F$. Since the Josephson frequency is given only by the potential difference across the junction regardless of the type of the potential, we take that this is the typical value for the maximal Josephson frequency also in the presence of spin-dependent potentials, which we now consider.

Using the estimated $\omega_J^{\max}=0.11E_F$, we show in Fig.~\ref{fig:asymplasma} the numerically obtained asymmetry in the plasma oscillation amplitudes given by Eq.~\eqref{eq:asymplasma} as a function of $\tilde{\delta}_{\downarrow}-\tilde{\delta}_{\uparrow}$ for different temperatures and various strengths of the attractive interaction in the typical regimes for an ultracold atom experiment. In particular, the used temperature regime is the same as in the experiment in~\cite{valtolina2015josephson} and the interaction strengths are in the same range as on BCS side of the BCS--BEC crossover in~\cite{valtolina2015josephson}. Note that we have used basic BCS equations in our calculations for simplicity, since we are interested only in the order of magnitude of the asymmetry. For the interaction strengths in Fig.~\ref{fig:asymplasma}, simple BCS theory estimates the critical temperature to be between $0.13T_F$ and $0.36T_F$. In more accurate schemes~\cite{zwerger2011bcs} there would be some corrections to the BCS parameter values. For example, the critical temperature is suppressed by a factor of roughly two. For a unitary Fermi gas, the critical temperature has been measured to be about $0.17T_F$~\cite{ku2012revealing}. We see in Fig.~\ref{fig:asymplasma} that the asymmetry grows for weaker interactions and can reach over 2\%. The asymmetry grows also with increasing temperature. 

\begin{figure*}[t!]
\centerline{
\includegraphics[scale=0.315]{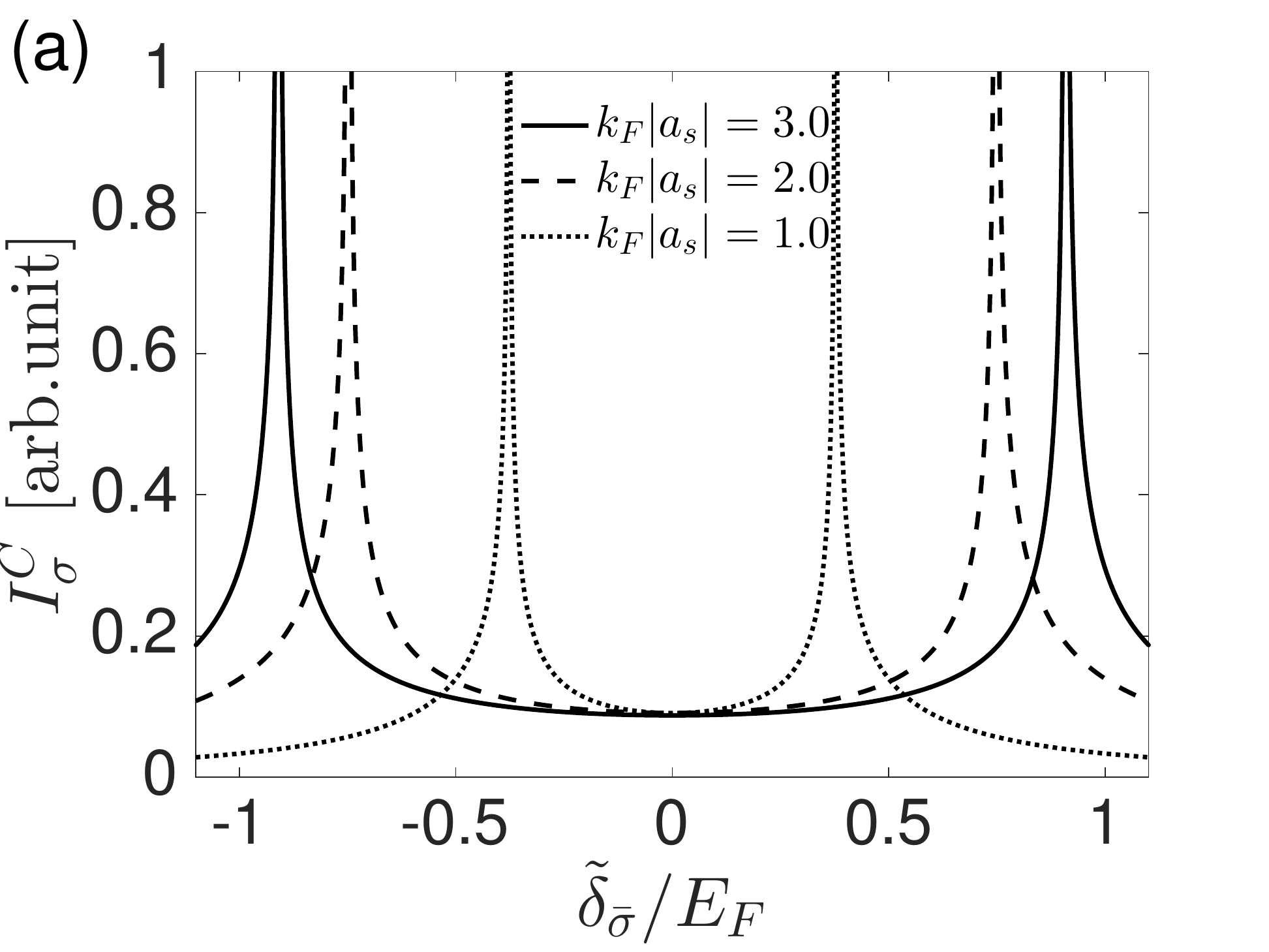}
\includegraphics[scale=0.315]{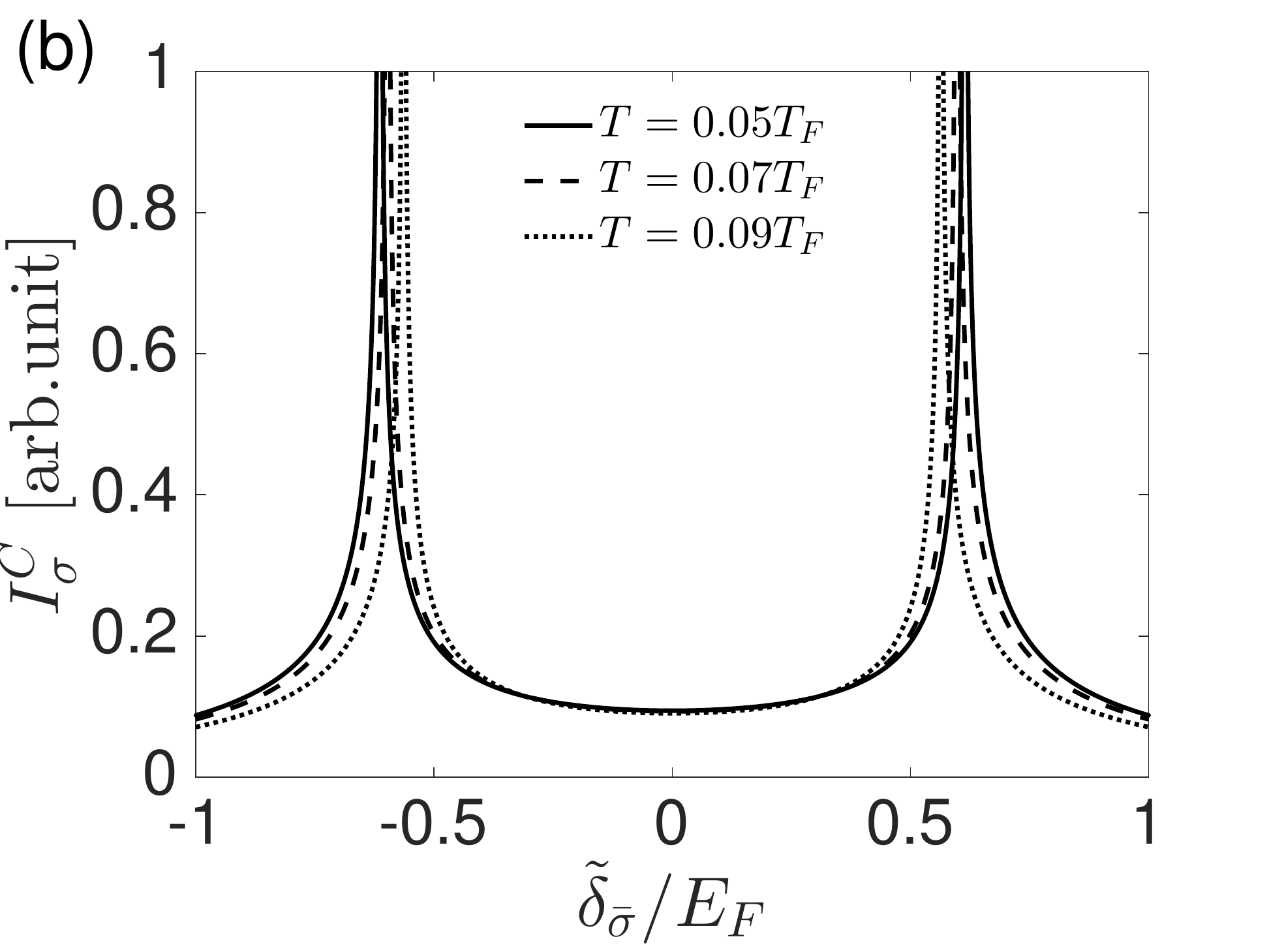}
}
\caption{Critical Josephson current $I^C_{\sigma}$ as a function of $\tilde{\delta}_{\bar{\sigma}}$ for (a) interaction $k_Fa_s=-3.0$ (solid curve), $k_Fa_s=-2.0$ (dashed curve), and $k_Fa_s=-1.0$ (dotted curve) at temperature $T=0.07T_F$, and (b) temperature $T=0.05T_F$ (solid curve), $T=0.07T_F$ (dashed curve), and $T=0.09T_F$ (dotted curve) for interaction $k_Fa_s=-1.5$. The divergence of $I^C_{\sigma}$ at $\tilde{\delta}_{\bar{\sigma}}=2\Delta$ is called the Riedel peak.}
\label{fig:critjose}
\end{figure*}

The behaviour of the asymmetry in the plasma oscillation amplitudes as a function of interaction strength and temperature is explained by the divergence of the critical Josephson current in Eq.~\eqref{eq:criticalc} at the Riedel peak at
$\tilde{\delta}_{\bar{\sigma}}=2\Delta$~\cite{riedel1964tunneleffekt}. The Riedel peak is located at a potential equal to the minimum energy required for creating a quasiparticle excitation, i.e., $2\min_{\bf k} E_{\bf k}=2\Delta$. The BCS quasiparticle density of states $D(E_{\bf k})$ then has a singularity at the corresponding quasiparticle energy $E_{\bf k}=\Delta$, i.e., at the gap edge. This singular behaviour of $D(E_{\bf k})$ is the physical reason behind the Riedel peak~\cite{PhysRev.147.255}. Since the gap $\Delta$ becomes smaller for weaker interactions and higher temperatures, the position of the Riedel peak moves closer to small frequencies, as shown in Fig.~\ref{fig:critjose}. Therefore, for decreasing interaction strength and increasing temperature it becomes easier to obtain greater asymmetries in the critical currents in the plasma oscillation regime and thus in the plasma oscillation amplitudes via Eq.~\eqref{eq:asymplasma}.

\begin{figure}[b!]
\centerline{
\includegraphics[scale=0.26]{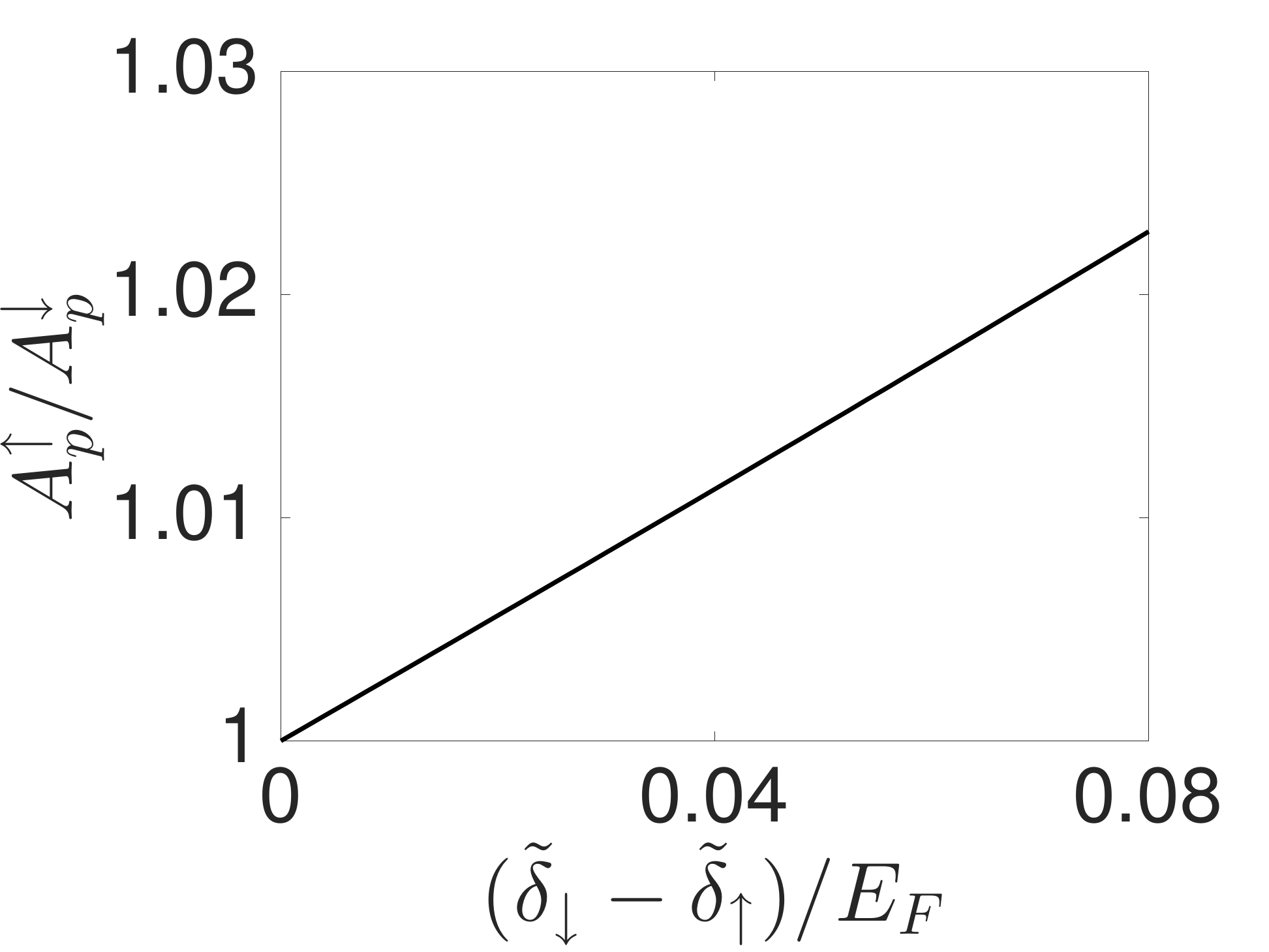}
}
\caption{Asymmetry in the spin-dependent plasma oscillation amplitudes as a function of the difference in the spin-dependent potentials for parameter values obtained in the experiment in~\cite{PhysRevLett.101.140403}. The Josephson frequency is $\omega_J=0.08E_F$, the interaction strength is $k_Fa_s=-4.0$, the temperature is $T=0.06T_F$, and the gap is $\Delta = 0.22E_F$.}
\label{fig:asymexp}
\end{figure}

We point out that since BCS theory overestimates the value of the superfluid gap~\cite{zwerger2011bcs}, the Riedel peak is actually closer to small frequencies than Fig.~\ref{fig:critjose} suggests. Therefore, in reality we can expect even greater asymmetries than those shown in Fig.~\ref{fig:asymplasma}. To demonstrate this, we plot in Fig.~\ref{fig:asymexp} the asymmetry in the plasma oscillation amplitudes using the experimental value for the gap, $\Delta = 0.22E_F$, obtained for interaction strength $k_F a_s = -4.0$ and temperature $T=0.06T_F$, as reported in~\cite{PhysRevLett.101.140403}. This experimentally determined gap is roughly half of the gap given by simple BCS theory. The corresponding maximal Josephson frequency is $\omega_J^{\max} \approx 0.08E_F$.  We see in Fig.~\ref{fig:asymexp} that even with this strong attractive interaction, asymmetries of over 2\% are feasible to obtain. Larger asymmetries can be expected for weaker interactions. 

We note that the amplitude asymmetry is always a fraction of an already small Josephson plasma mode signal ($z=3\%$ in~\cite{valtolina2015josephson}). This makes the detection of the asymmetry challenging. However, the Josephson plasma mode is a collective dipole oscillation (i.e., a center-of-mass motion of all atoms), and its amplitude can be detected in a time-of-flight (TOF) expansion. Thus, the evolution of the population imbalance is mapped onto the center-of-mass displacement. For the parameter values in~\cite{valtolina2015josephson}, a displacement of several tens of micrometers can be achieved with a short TOF of duration less than 10 ms. This significantly increases the signal-to-noise ratio for detecting the asymmetry. 

Finally, we make a comparison to spin diffusion. The Josephson plasma oscillation occurs on a time scale on the order of the trapping frequency, i.e., 50 ms in~\cite{valtolina2015josephson}, while longitudinal spin diffusion takes place on a much longer time scale (200 ms with the parameters of~\cite{valtolina2015josephson}). This means that while the plasma oscillation is happening, the spin bias does not change significantly. This implies that one can properly define a chemical potential, the charging energy, and also the spin-dependent potential $\delta_\sigma$. In other words, the plasma oscillation occurs in a quasi-equilibrium configuration. In addition, for measuring the spin-asymmetric plasma oscillation amplitude, only half of the trapping period is required since we only need the height from the maximum peak to the minimum one. For these reasons, we do not expect spin diffusion to significantly affect the possible observation of the spin-asymmetric plasma oscillations.

\section{Summary and discussion}\label{sec:summary}

In summary, we have studied the plasma oscillation regime in a Josephson junction with a spin-dependent potential, realized with an ultracold superfluid Fermi gas. We have proposed methods to experimentally create the required spin-dependent potential across the junction. We have predicted that in this setup the Josephson plasma oscillation amplitude becomes spin-dependent but the plasma frequency is the same for both spin-components similarly to the full ac spin-asymmetric effect. The spin-asymmetry in the plasma oscillation amplitudes is given by the asymmetry in the spin-dependent critical Josephson currents which are characteristic of the spin-asymmetric Josephson effect. Furthermore, we have shown that the asymmetry in the amplitudes can be tuned by varying the spin-dependent potentials. In the parameter regime typical of ultracold atom experiments, we have demonstrated that asymmetries on the order of a couple of per cent are achievable. The observation of these spin-dependent plasma oscillations would establish the so far undetected spin-asymmetric Josephson effect. 

As a remark, we mention that an effective bosonic picture is often adopted when the usual Josephson phenomenon is described. We want to point out that the spin-asymmetric Josephson effect shows that this bosonic description is incomplete and underlines the fact that the contribution from the fermionic single particles that compose the Cooper pairs cannot be overlooked, as manifested by the previously unnoticed single-particle interference processes in the Josephson current. Importantly, the contribution from the single-particle processes leads to a measurable spin-polarized supercurrent. Noteworthily these single-particle interferences are present also in the standard spin-symmetric case, but there their effect may be difficult to detect.  

Finally, we state that, as shown in~\cite{miikkajose,kreula2014}, the spin-asymmetric Josephson effect and its direct-current limit, i.e., $\omega_J=0$, provide a critical supercurrent which can be tuned with the spin-dependent potentials. This feature has the potential be useful in a variety of technological applications if realized in solid-state systems.

\section*{Acknowledgements}

We thank  Matteo Zaccanti, Giacomo Roati, and Dieter Jaksch for useful discussions. This work
was supported by the Academy of Finland through its Centres of Excellence Programme and under Project No. 284621, and by the European Research Council (ERC-2013-AdG-340748-CODE). JMK acknowledges financial support from Christ Church, Oxford and the Osk.\ Huttunen Foundation. GV was supported by the European Research Council Grant No.\ 307032 QuFerm2D.

\section*{Appendix: Linear response and Kadanoff--Baym formalism}

\renewcommand{\theequation}{A\arabic{equation}}
\setcounter{equation}{0}

Here, we elucidate some of the details in the derivation of Eq.~\eqref{eq:josecurr}.
First, note that we split the Hamiltonian in Eq.~\eqref{eq:totalham} of the main text into three parts. The reason for this is that we calculate the current treating $\hat{H}_{\Omega}$ as the first order perturbation in linear response theory while separating $\hat{H}_{\delta}$ from $\hat{H}_{0}$ allows us to take two simple BCS superfluids as the unperturbed initial state. The Josephson current for spin $\sigma$ is then given by
\begin{align}\label{eq:joseaux}
I^J_{\sigma}(t)=&-2\, {\rm Im}\Big[e^{-i\left (\tilde{\delta}_\uparrow + \tilde{\delta}_\downarrow\right)t } \Omega_\uparrow \Omega_\downarrow \nonumber \\ & \times L\big( {\bf p=0}, -\tilde{\delta}_{\bar{\sigma}}+i0^+ \big)  \Big],
\end{align}
in which $\tilde{\delta}_\sigma=\mu_L-\mu_R-\delta_\sigma$ and $L\big( {\bf p}, -\tilde{\delta}_{\bar{\sigma}}+i 0^+)$ is the Fourier transform of the retarded linear response function which is defined in position space and time domain as
\begin{align}
&L({\bf r_1}t,{\bf r_3}t,{\bf r_2}t',{\bf r_4}t')=-i\theta(t-t') \nonumber \\ & \times \left \langle \left [\hat{\psi}^{\dagger}_{\uparrow,R}({\bf r_3}t)\hat{\psi}_{\uparrow,L}({\bf r_1}t),\hat{\psi}^{\dagger}_{\downarrow,R}({\bf r_4}t')\hat{\psi}_{\downarrow,L}({\bf r_2}t') \right ] \right \rangle,
\end{align}
where $\theta(t)$ is the Heaviside step function, $[\cdot, \cdot]$ denotes the commutator, and the thermodynamic average is calculated with respect to the unperturbed Hamiltonian $\hat{H}_0$. Further, the time-dependent field operators are in the interaction picture, with ${\hat{\psi}^{\dagger}_{\sigma,L/R}({\bf r},t)=e^{i\hat{H}_0t}\hat{\psi}^{\dagger}_{\sigma,L/R}({\bf r})e^{-i\hat{H}_0t}}$.

To get the Josephson current, we need to obtain the linear response function $L$ in Eq.~\eqref{eq:joseaux}. As explained in~\cite{miikkajose}, using the Kadanoff--Baym method~\cite{baym1961conservation,kadanoff1962quantum} to obtain the four-operator correlator $L$ as a variational derivative of the single-particle Green function $G$ with respect to the coupling $\Omega$, i.e., $L=-\frac{\delta G}{\delta \Omega}|_{\Omega=0}$, ensures that $L$ obeys the same conservation laws as $G$, and thus the linear response calculations are self-consistent. For details, see Refs.~\cite{baym1961conservation,kadanoff1962quantum}.  However, in practice the analytic form of $G(\Omega)$ is not known, and thus taking the variational derivative is not possible. Instead, one can derive an expression for $L$ using equations of motion for the Green function. In what follows, we present the general idea of the Kadanoff--Baym approach to obtain $L$ for completeness and use a shorthand notation for all position and time variables as per Refs.~\cite{baym1961conservation,kadanoff1962quantum}.

We proceed to derive an integral equation from which $L$ can be solved. We start from the relation $\int d\bar{1}\, G(1,\bar{1})G^{-1}(\bar{1},1')=\delta(1-1')$, from which one obtains the variational derivative of $G$ with respect to $\Omega$ as
\begin{align}\label{eq:auxG}
\frac{\delta G(1,1')}{\delta \Omega(2',2)}=- \int d\bar{3} \, d\bar{4} \, G(1,\bar{3}) \frac{\delta G^{-1}(\bar{3},\bar{4})}{\delta \Omega(2',2)}G(\bar{4},1').
\end{align}
The left hand side of this equation becomes the linear response function $L(12,1'2')$ when evaluated at $\Omega=0$ and multiplied by -1. We modify the right hand side of Eq.~\eqref{eq:auxG} so that the linear response function appears there as well. For this, we use the equation of motion for the Green function, given by~\cite{baym1961conservation}
\begin{align}
G^{-1}(\bar{3},\bar{4})=G_0^{-1}(\bar{3},\bar{4})-\Omega(\bar{3},\bar{4})-\Sigma(\bar{3},\bar{4}),
\end{align}
where $G_0$ is the non-interacting Green function which does not depend on $\Omega$, and $\Sigma$ is the self-energy which is a functional of $G$ and $\Omega$. Differentiating this equation and using the chain rule for differentiation for the self-energy, namely
\begin{align}
\frac{\delta \Sigma(\bar{3},\bar{4})}{\delta \Omega(2',2)}=\int d\bar{5} \, d\bar{6} \, \frac{\delta \Sigma(\bar{3},\bar{4})}{\delta G(\bar{5},\bar{6})} \frac{\delta G(\bar{5},\bar{6})}{\delta \Omega(2',2)},
\end{align}
directly yield an integral equation for $L$ as
\begin{align}\label{eq:Leq}
&L(12,1'2')=-G(1,2')G(2,1') \nonumber \\ & + \int d\bar{3} \, d\bar{4} \, G(1,\bar{3}) G(\bar{4},1') \frac{\delta \Sigma(\bar{3},\bar{4})}{\delta G(\bar{5},\bar{6})} L(\bar{5} 2,\bar{6} 2'),
\end{align}
where all quantities on the right hand side are evaluated at ${\Omega=0}$. From Eq.~\eqref{eq:Leq} the linear response function can be solved at least numerically in the general case. However,  in our system we are able to continue analytically.

As is evident in Eq.~\eqref{eq:joseaux}, we require $L$ as a function of momentum and frequency. Thus, we Fourier-transform Eq.~\eqref{eq:Leq} into momentum and frequency space. We are then left with a matrix equation of the form $L=\Pi + M L$, where $\Pi$ and $M$ are matrices coming from the first and second term of Eq.~\eqref{eq:Leq}, respectively. From this matrix equation $L$ can be solved analytically in some special cases. For example, in the case of BCS mean-field theory and a contact interaction as considered here and in~\cite{miikkajose}, a closed form for the derivative of the self-energy can be obtained. The required linear response function then becomes~\cite{miikkajose}
\begin{align}\label{eq:analL}
L({\bf p},i\omega_n)=-\Pi_{\mathcal{F}} ({\bf p}, i\omega_n ),
\end{align}
where $\Pi_{\mathcal{F}} ({\bf p}, i\omega_n )$ is given by Eq.~\eqref{eq:Fbubble}. Applying the analytic continuation from Matsubara to real frequencies, ${i\omega_n \rightarrow {-\tilde{\delta}_{\bar{\sigma}}+i0^+}}$, to $L$ in Eq.~\eqref{eq:analL} yields Eq.~\eqref{eq:josecurr}.

To obtain a value for the critical Josephson current, the energy gap $\Delta$ in the anomalous Green function in Eq.~\eqref{eq:anomalousG} is obtained numerically by solving the BCS gap equation (see, e.g.,~\cite{stoof2009ultracold,ueda2010fundamentals})
\begin{align}
\frac{m}{4\pi |a_s|}=\frac{1}{2V}\sum_{\bf k}\left( \frac{1-2n_F(E_{\bf k})}{E_{\bf k}} - \frac{1}{\epsilon_{\bf k}} \right),
\end{align}
where  $n_F(E_{\bf k})=1/\left[\exp(\beta E_{\bf k})+1 \right]$, along with the number equation 
\begin{align}
n=\frac{1}{V}\sum_{\bf k} \left\{1- \frac{\xi_{\bf k}}{E_{\bf k}}\left[1-2n_F(E_{\bf k})\right] \right\}.
\end{align}
We assume for simplicity that the superfluid gap is unaffected by the presence of the spin-dependent potentials.


%

\end{document}